\documentclass[modern]{aastex631}

\accepted{February 18, 2022}

\submitjournal{ApJ}

\shorttitle{Wakefield Acceleration in a jet from a NDAF}
\shortauthors{Kato et al.}
\graphicspath{{./}{figures/}}

\usepackage{amsmath}

\def\lsim{\lower 2pt \hbox{$\, \buildrel {\scriptstyle <}\over
{\scriptstyle \sim}\,$}}
\def\gsim{\lower 2pt \hbox{$\, \buildrel {\scriptstyle >}\over
{\scriptstyle \sim}\,$}}

\begin{document}

\title{Wakefield Acceleration in a Jet from a Neutrino Driven Accretion Flow around a Black Hole}

\correspondingauthor{Yoshiaki Kato}
\email{yoshiaki.kato@riken.jp}

\author[0000-0003-2349-9003]{Yoshiaki Kato}
\affiliation{RIKEN \\
2-1 Hirosawa, Wako\\
Saitama, 351-0198, Japan}

\author[0000-0002-3918-1166]{Toshikazu Ebisuzaki}
\affiliation{RIKEN \\
2-1 Hirosawa, Wako\\
Saitama, 351-0198, Japan}

\author{Toshiki Tajima}
\affiliation{Department of Physics and Astronomy, University of California\\
Irvine, CA 92679, USA}

\begin{abstract}

We have investigated electro-magnetic (EM) wave pulses in a jet from a neutrino driven accretion flow (NDAF) around a black hole (BH).
NDAFs are massive accretion disks whose accretion rates of $\dot{M}\approx 0.01 - 10 \mathrm{M}_\odot/\mathrm{s}$ for stellar-mass BHs.
Such an extreme accretion may produce a collimated relativistic outflow like a magnetically driven jet in active galactic nuclei and micro-quasars.
When we consider strong toroidal magnetic field stranded in the inner-region of a NDAF disk and magnetic impulses on the jet, we find that they lead to the emanation of high energy emissions for gamma-ray bursts as well as high energy cosmic rays.
When Alfv\'enic wave pulses are generated by episodic immense accretions, it propagates along the large-scale structured magnetic field in the jet.
Once the Alfv\'enic wave pulses reach at nearly the speed of light in the underdense condition, it turns into EM wave pulses which produce plasma wakes behind them.
These wakefields exert a collective accelerating force synchronous to the motion of particles.
As a result, the wakefield acceleration premises various observational signatures, such as pulsating bursts of high energy gamma-rays from accelerated electrons, pulses of neutrinos from accelerated protons, and protons with maximum energies beyond $10^{20}~\mathrm{eV}$.

\end{abstract}



\section{Introduction} \label{sec:intro}

Gamma-ray bursts \citep[GRBs;][]{1973ApJ...182L..85K} are sudden ($\sim$ seconds) flashes of gamma-rays with energy $\sim100~\mathrm{keV}$ that arrive at the Earth several times a day \citep[e.g.,][]{1995ARA&A..33..415F}. 
GRBs are classified into two categories by its duration: one is that long GRBs last longer than $2$ seconds and usually consist of many pulses in which the total energy of gamma-ray emissions is about $10^{51}~\mathrm{erg}$ \citep{2006ARA&A..44..507W} after correction of the relativistic beaming effect, and the other is that short GRBs last less than $2$ seconds and consist of one pulse with the beaming-corrected total energy of about $10^{50}~\mathrm{erg}$ \citep{2015ApJ...815..102F}.
The short GRBs are speculated to be mergers of two compact objects, either two neutron stars (NSs) or NS - black hole (BH) binaries.
In merging a binary of neutron stars (BNS), the tidal force of the more massive NS destroys the other, and therefore debris of less massive NS forms a massive accretion disk.
This speculation is beautifully proved by the observation of an association between a short GRB (GRB 170817A) and the gravitational wave burst detected by LIGO and VIRGO \citep{2017ApJ...850L..35A}.
The long GRBs on the other hand had been surmised to be associated with the core-collapse of a massive star for years.
This conjecture was established by the discovery of GRB on April 1998 (GRB 980425), which has a connection with an unusual supernova 1998bw \citep{1998Natur.395..670G}.
The successive studies of supernovae (SNe) accompanied by GRBs suggest that long GRBs are associated with SNe of type Ic that have $30 - 50$ times more  energetic than normal SNe, the so-called hypernovae \citep[HNe;][]{1998Natur.395..672I,2013ARA&A..51..457N}.
Such extreme explosion energy of HNe cannot be explained by the conventional theory of SNe.
This is one of the major motivation of our study.
Meanwhile \citet{1993ApJ...405..273W} proposed "collapsar model" of GRBs in which the entire mass in the core cannot fall down directly to a newly born black hole (BH) or neutron star (NS) but form an accretion disk, if the specific angular momentum of the core is higher than a critical value of $2\sqrt{3}G/c\sim1.5\times 10^{16} ~\mathrm{cm}^{2}\,\mathrm{s}^{-1}$.
The accretion disk is powered for a longer period of time by the collapsing star and radiates thermal emission via viscous dissipation \citep{shakura+sunyaev:1973AA}.
For hyper-critical accretion rates, $\dot{M} \gg L_\mathrm{Edd}/c^{2}$ where $L_\mathrm{Edd}$ is the Eddington luminosity \citep{1979rpa..book.....R}, optical depth becomes too high for photons to escape, and therefore the neutrino cooling takes over the radiation cooling \citep{1992ApJ...395L..83N}.
This regime is what we referred to as a neutrino driven accretion flow (NDAF) \footnote{Generally, NDAF is the abbreviation of neutrino-dominated accretion flow.}. 
\citet{1999ApJ...524..262M} performed hydrodynamic simulations of the core-collapse.
The resultant disk turns out to be thick against neutrino interaction and is evolved into a NDAF disk.
They showed the neutrino annihilation at the rotation axis of the disk produces a fireball to launch bipolar mildly relativistic jets.
They also discuss that the magnetohydrodynamical (MHD) process might produce bipolar jets in a similar manner to the neutrino annihilation, if magnetic energy dissipation took place in and above the disk.
Since the kinetic luminosity of the jets is a few $10^{51}~\mathrm{erg}\,\mathrm{s}^{-1}$, they speculated the jets can produce GRBs.
The pioneering magneto-hydrodynamic (MHD) simulations of magnetically driven jets from accretion disks were performed by \citet{1985Natur.317..699U} and \citet{1985PASJ...37...31S,1986PASJ...38..631S}.
They found that the jets are accelerated by torsional Alfv\'en waves propagating along magnetic field lines anchored to the disks \citep[the so-called "sweeping magnetic twist mechanism"; see also][]{2001Sci...291...84M}.
This mechanism requires the strong poloidal magnetic field structure above the disk \citep[also known as "beads on wires mechanism";][]{1982MNRAS.199..883B}.
However, \citet{1990ApJ...350..295S} postulated another kind of magnetically driven jet, in which toroidal magnetic fields are dominant in the jet \citep[see also][]{1990PASJ...42..793F}.
Such toroidal magnetic fields can be generated inside the accretion disk as a result of magneto-rotational instability \citep[MRI;][]{1991ApJ...376..214B}.
In other words, the presence of magnetic fields in the accretion disks is not only the source of turbulence and viscosity causing structure deformation but also the driver of mega-parsec-scale structure formation such as astrophysical jets, radio lobes, and cocoons, the largest structures in the Universe \citep{1996A&ARv...7....1C,1997ARA&A..35..607Z, 2014SSRv..183..405H,2016A&ARv..24...10T}.
Through these examples, one obtains the notion of the magnetic fields in the Universe as a stimulus or catalyst of the Universe's structure formation.
This idea was demonstrated for the first time by \citet{2002PASJ...54..121K}.
Later, \citet{2004ApJ...605..307K} have found the formation of self-collimated magnetic field structure emanating from the magnetized accretion disk \citep[so-called "magnetic tower";][]{1996MNRAS.279..389L}.
For collapsar model, the emergence of magnetic tower from the NDAFs is promising \citep{2007ApJ...669..546U} because large-scale toroidal magnetic field of $B=10^{15}~\mathrm{G}$ or beyond can be generated in a rapidly rotating core-collapsing star \citep{2003ApJ...584..954A,2015Natur.528..376M}.
According to \citet{2013ARA&A..51..457N}, there are two branches of supernovae whose progenitors are main-sequence stars with the mass of $20 - 25~\mathrm{M}_\odot$ depending on its angular momentum: one is an energetic bright "hypernova" branch for a fast-rotation core, and the other is a faint, low-energy "failed supernova" branch for a slow-rotation core.
It has been suggested that the difference between the former and the latter is caused by MHD processes mentioned above.
%

In both short and long GRBs, the central engine is NDAF disks.
\citet{1999ApJ...518..356P} investigated the energy extraction from both NDAFs and rotating BHs.
Such models have been extensively developed by many groups \citep{2002ApJ...577..311K,2002ApJ...579..706D,2004MNRAS.355..950J,2005ApJ...632..421L,2007ApJ...657..383C, 2013ApJ...766...31K}.
Theoretical progress and applications of NDAF disks have been summarized in recent review by \citet{2017NewAR..79....1L}.
Under the assumption that the BH spin is moderately fast and the magnetic pressure near the horizon is limited by the inner disk pressure of NDAFs, it has been conjectured that Blandford-Znajek (BZ) mechanism \citep{1977MNRAS.179..433B} exceed the energy deposition rate expected from neutrino pair-annihilation above the NDAF.
However, the neutrino pair-annihilation is not the only mechanism for extracting the energy from the NDAF disks and, more importantly, MHD processes other than the BZ mechanism may play an important role.
For example, the presence of magnetic fields in the NDAF disks could also explain repeatable short-duration variability in long GRBs in the same manner as strong variability of gamma-rays in blazers \citep{2020MNRAS.493.2229C}.
Another point of the supporting evidence of the magnetic fields in the accretion disk and its jets triggering electron acceleration may be seen in the NS - NS collision triggered gamma-ray burst emission \citep{Abbott:2017it}, which accompanied simultaneous gravitational wave emission.
Such gamma-ray emission was predicted via wakefield acceleration by \citet{Takahashi2000} \citep[see also][]{2002PhRvL..89p1101C}.
Moreover, BHs may not be necessary because winds and outflows from either rapidly spinning magnetars or NDAF disks could be a central engine of long GRBs and HNe \citep{2015MNRAS.451..282S,2019ApJ...871..117S}.
Not to mention that the system of a BH/NS and a massive accretion disk show a huge amount of energy emissions via a variety of channels, such as radio-waves, infrared and optical emissions, X-rays and gamma-rays, ultra high energy cosmic rays (UHECR), neutrinos, and even gravitational waves.
Finally, even if the appropriate system of a BH/NS and an accretion disk is given, the greatest uncertainty is the acceleration mechanism of high energy particles that are responsible for the production of high energy emissions in the jets.
We will propose an alternative mechanism for the origin of high-energy emissions.
Recently, \citet{ebisuzaki+tasjima2019} have proposed a model of acceleration of charged particles to very high energies including energies above $10^{20}~\mathrm{eV}$ for  the case of protons and nucleus, and $10^{12-15}~\mathrm{eV}$ for electrons by electro-magnetic wave-particle interaction.
If the episodic eruptive accretions generate Alfv\'enic wave pulses along the magnetic field in the jets, such Alfv\'enic wave pulses act as a driver of the collective accelerating pondermotive force whose direction is parallel to the motion of particles.
This pondermotive force drives the wakes.
Because the wakes propagate at the same speed with the particles, the so-called wakefield acceleration has a robust built-in coherence in the acceleration system itself.
In other words, the accelerating particles are surfing along with the wakes without too much energy loss due to synchrotron emission during the process.
This reinforces the blazer emission mechanism of gamma-ray photons by this mechanism as a natural way.
It is therefore the wakefield acceleration have more advantage than the diffusive shock accelerations \citep{Drury_1983}.
%

\begin{figure}
\epsscale{0.85}
\plotone{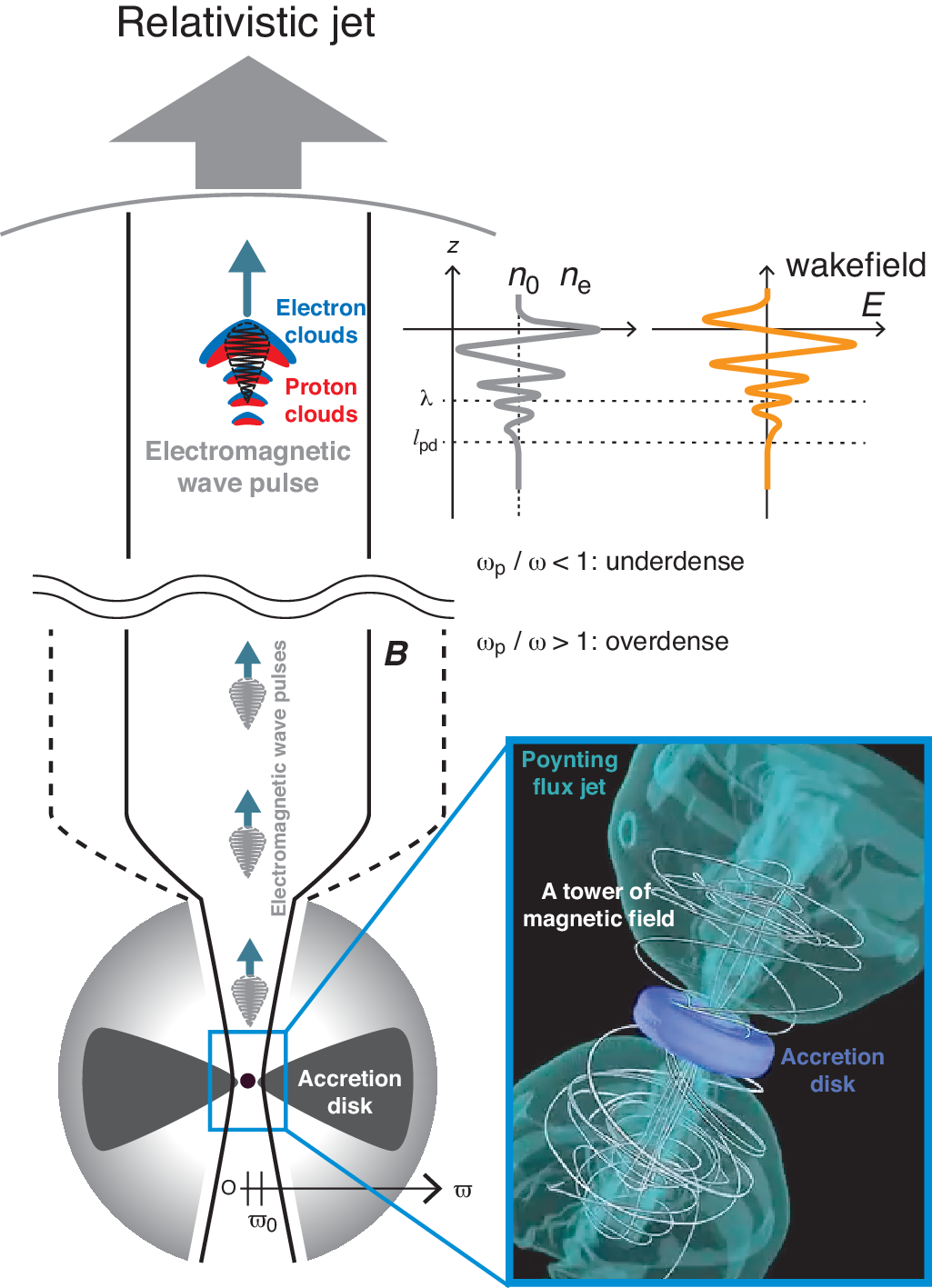}
\caption{Schematic illustration of a neutrino driven accretion flow (NDAF) and a magnetically driven jet.  The jet is driven by self-collimated magnetic field structure emanating from a magnetized accretion disk \citep{2004ApJ...605..307K,Kato:2006fu}.  Electro-magnetic (EM) wave pulses are generated by magnetohydrodynamic instabilities in the magnetized disk and are propagating along the jet in the presence of magnetic fields in the jet plasma \citep{1973bppp.book.....I,1981ASSL...82.....A,1997plas.conf.....T}, even though the frequency of EM wave pulses $\omega$ is smaller than the plasma frequency $\omega_\mathrm{p}$ in the vicinity of NDAFs.  If the frequency of EM wave pulses becomes larger than the plasma frequency ($\omega > \omega_\mathrm{p}$) in the distance, the energy of EM wave pulses is transferred into charged particles as a result of the wakefield acceleration \citep{1979PhRvL..43..267T,ebisuzaki+tasjima2019}.}
\label{fig:schematic}
\end{figure}

In this study we extend a model of the accretion disk presented by \citet{ebisuzaki+tasjima2019} \citep[see also][]{Ebisuzaki:2014es,2018MNRAS.479.2534M} into a NDAF.
A basic concept of our model is shown as a schematic illustration in Figure~\ref{fig:schematic}.
We estimate the energy flux of both the electro-magnetic wave pulses and neutrino emissions from the NDAFs.
Section \ref{sec:model} we describe more details how we derived the structure of NDAFs for a reader's convenience although the basic equations we used here are basically same as the previous study.
We present our results in Section \ref{sec:results1} and  \ref{sec:results2}.
We summarize in Section \ref{sec:summary_and_discussion} that observational signature of the wakefield acceleration in relation to GRBs with some discussions, and we conclude in Section \ref{sec:conclusion}.
%

\section{Model}\label{sec:model}

\subsection{Neutrino driven accretion flow disks}\label{sec:model:disk_model}
We assume that an axisymmetric steady-state accretion disk is in Keplerian rotation in which the radial ($\varpi$-) distribution of orbital angular velocity $\Omega_\mathrm{K}(\varpi)\equiv \sqrt{GM/\varpi^{3}}$ under the Newtonian gravity potential where $M$ is the mass of a central BH and $G$ is the gravitational constant.
Under the hydrostatic equilibrium in the vertical ($z$-) direction, the vertical structure is expressed to be the Gaussian profile with the pressure scale height $H(\varpi)$ where $\varpi$ is the distance from the gravity center on the mid plane where $z=0$.
It  is therefore the basic equations are formulated by using vertically integrated  hydrodynamic variables of the disk, such as the column density $\Sigma(\varpi)\equiv\int_{-\infty}^{\infty}\rho(\varpi,z) dz$ where $\rho(\varpi,z)$ is the density distribution in the $\varpi$-$z$ plane.
In the one-zone approximation of the vertical structure of the disk, the radial density distribution at $z=0$ is expressed as $\rho_\mathrm{0}(\varpi) = \Sigma(\varpi)/2H(\varpi)$.
%

For a constant mass accretion rate $\dot{M}$ though the disk in steady-state solution, the mass conservation equation in the radial direction of the disk yields:
\begin{equation}
\dot{M}=-2\pi \varpi\Sigma(\varpi) v_\varpi(\varpi)=\mathrm{const.},
\label{conservation_of_mass}
\end{equation}
where $v_\varpi$ is the radial velocity of the gas in the disk, which is negative for inflow.
Here we consider the mass accretion rate is extremely high as $\dot{M} \geq  0.1 \mathrm{M}_\odot/s$.
%

From angular momentum conservation in the steady-state disk, we obtain a formula of angular momentum transport rate:
\begin{equation}
\dot{M} \varpi^{2} \Omega_\mathrm{K}(\varpi)=-2\pi \varpi^{2}\cal{S}_{\varpi\varphi} + \mathrm{const.},
\label{conservation_of_angular_momentum}
\end{equation}
where ${\cal S}_{\varpi\varphi} = \alpha\Sigma(\varpi) {\cal C}_\mathrm{s}(\varpi)^{2}$ is the viscous stress given in the $\alpha$-disk prescription \citep{shakura+sunyaev:1973AA}, where ${\cal C}_\mathrm{s}(\varpi)$ is the sound velocity.
We set a constant viscosity parameter of $\alpha = 0.1$.
%

Under the one-zone approximation in the isothermal condition, the hydrostatic balance in the vertical direction is reduced to $H(\varpi)={\cal C}_\mathrm{s}(\varpi)/\Omega_\mathrm{K}(\varpi)$.
In the case of a torque-free boundary condition ${\cal S}_{\varpi\varphi}=0$ at the inner edge of the disk $\varpi=\varpi_\mathrm{in}$, Equation\,\ref{conservation_of_angular_momentum}  becomes
\begin{equation}
\dot{M} \varpi^{2}_\mathrm{in} \Omega_\mathrm{K} (\varpi_\mathrm{in}) = \mathrm{const.} \equiv \cal{L}_\mathrm{in}
\label{conservation_of_angular_momentum_at_rin}
\end{equation}
where $\cal{L}_\mathrm{in}$ is the angular momentum gain of the central BH.
But, since we are not interested in the evolution of a spin of a BH, we simply ignore the angular momentum gain, ${\cal L}_\mathrm{in} = 0$.
From Equation \ref{conservation_of_angular_momentum},  we have
\begin{equation}
\Sigma(\varpi) = \frac{\dot{M}\Omega_\mathrm{K}(\varpi)}{2\pi\alpha{\cal C}_\mathrm{s}^{2}(\varpi)}.
\label{conservation_of_angular_momentum2}
\end{equation}

\begin{figure}
\epsscale{0.85}
\plotone{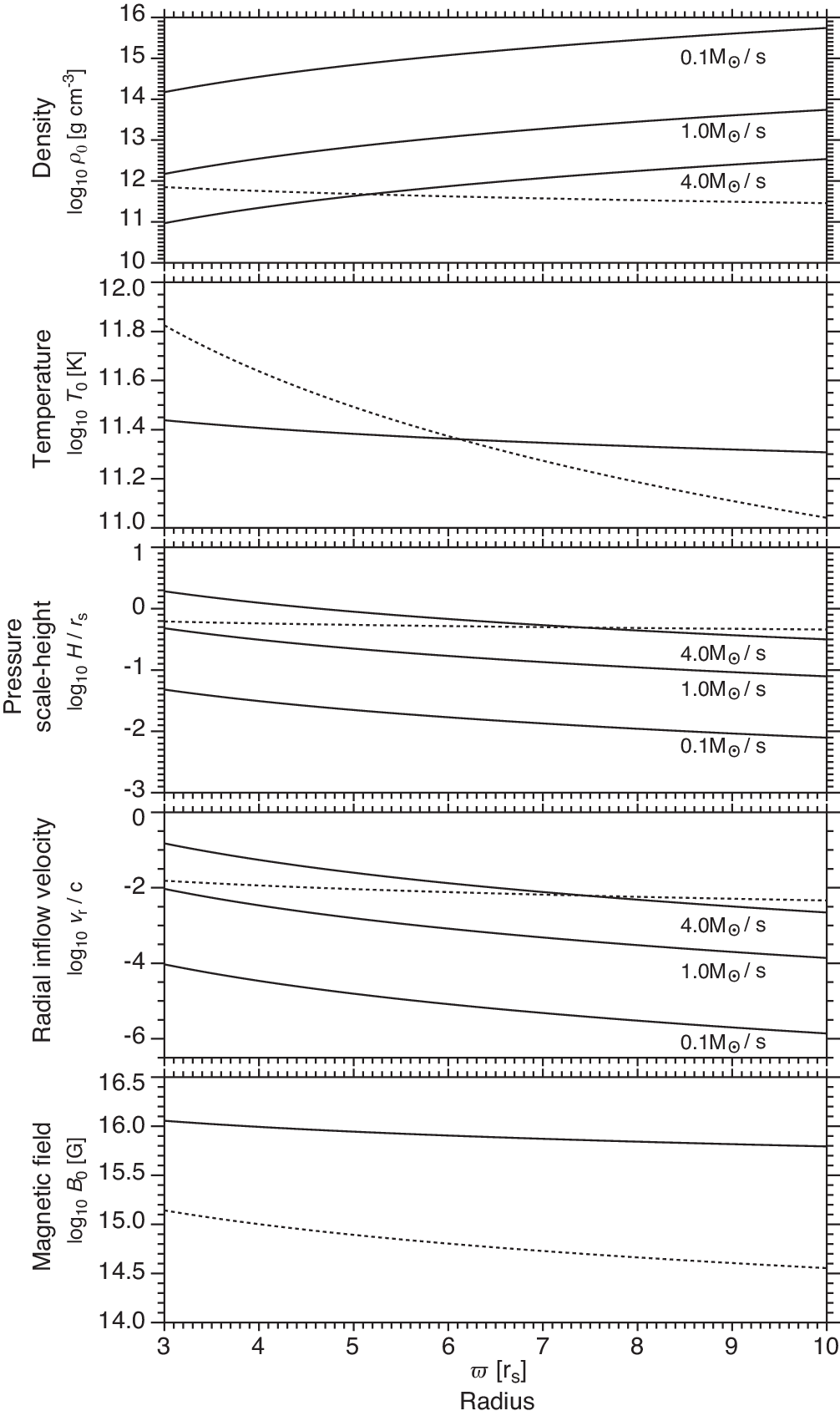}
\caption{Analytical solution of an accretion disk under the condition of local radiative equilibrium by neutrino emission with mass accretion rates of $\dot{M}/\dot{\mathrm{M}}_{\odot}=0.1$, $1.0$, and $4.0$.  Temperature ($T_\mathrm{0}$) and magnetic field ($B_\mathrm{0}$) are independent of $\dot{M}$. $B_\mathrm{0}$ reaches $10^{16}\,\mathrm{G}$ at the inner-edge of the disk.  The value of physical properties at the inner region of the NDAF disk is almost consistent with that of \citet{2013ApJ...766...31K} for $\dot{M}/\dot{\mathrm{M}}_{\odot}=1$ as shown in the dotted curves.}
\label{fig:diskmodels}
\end{figure}

In the standard accretion disk model \citep{shakura+sunyaev:1973AA}, the viscous heating rate $Q_\mathrm{visc}$ is determined as
\begin{equation}
Q_\mathrm{vis}(\varpi) = \frac{3\dot{M}}{4\pi}\Omega_\mathrm{K}^{2}(\varpi).
\end{equation}
For a steady-state accretion flow, the viscous heating and neutrino cooling must be balanced at every radius, $Q_\mathrm{vis}(\varpi)=Q_\nu(\varpi)$, where $Q_\nu(\varpi)$ is the neutrino energy cooling rate.
Since the neutrino can escape from the upper and lower side of the disk,  so the emergent neutrino energy flux is written to: 
\begin{equation} 
{\cal F}_{\nu}(\varpi) = Q_{\nu}(\varpi) / 2 =  \frac{3\dot{M}}{8\pi}\Omega_\mathrm{K}^{2}(\varpi).
\label{eqn:radiation_flux}
\end{equation}
The total energy density at the surface density $\Sigma(\varpi)$ is determined by the relation \citep{shakura+sunyaev:1973AA},
\begin{equation} 
\epsilon_\mathrm{0}(\varpi)=\frac{3}{4}\frac{{\cal F}_{\nu}(\varpi)}{c}\bar{\kappa}_{\nu}(\varpi)\Sigma(\varpi) = \frac{9\dot{M}}{32\pi c}\bar{\kappa}_{\nu}(\varpi)\Sigma(\varpi)\Omega_\mathrm{K}^{2}(\varpi)
\end{equation}
where $c$ is the speed of light and $\bar{\kappa}_{\nu}(\varpi)$ is the Rosseland mean opacity of neutrino.
In the neutrino driven accretion flow,  the sound speed is expressed as
\begin{equation} 
{\cal C}_\mathrm{s}(\varpi) = \sqrt{\frac{\epsilon_\mathrm{0}(\varpi)}{3\rho_\mathrm{0}(\varpi)}}
\label{eqn:sound_speed}
\end{equation}
where $\rho_\mathrm{0}(\varpi)$ is the gas density  on the mid-plane in the disk.
Substituting Equation \ref{eqn:sound_speed} into the relation of the hydrostatic balance in the vertical direction, we find that the pressure scale height becomes
\begin{equation}
H(\varpi) = \frac{3\dot{M}}{16\pi c}\bar{\kappa}_\nu(\varpi)
\label{eqn:scaleheight}
\end{equation}
Using the above equations, the properties of a neutrino driven accretion flow are summarized as follows:
\begin{equation}
\rho_\mathrm{0}(\varpi) = \frac{1024\pi^{2}c^{3}}{27\alpha\dot{M}^{2}}\bar{\kappa}^{-3}_\nu(\varpi)\Omega^{-1}_\mathrm{K}(\varpi).
\label{eqn:ro0}
\end{equation}
\begin{equation}
\epsilon_\mathrm{0}(\varpi) = \frac{4c}{\alpha}\bar{\kappa}^{-1}_{\nu}(\varpi)\Omega_\mathrm{K}(\varpi).
\label{eqn:epsilon0}
\end{equation}
The pressure of the disk $p_\mathrm{0}(\varpi) = \rho_\mathrm{0} {\cal C}_\mathrm{s}^{2}(\varpi) = \epsilon_\mathrm{0}(\varpi) / 3$ becomes
\begin{equation}
p_\mathrm{0}(\varpi) = \frac{4c}{3\alpha}\bar{\kappa}^{-1}_{\nu}(\varpi)\Omega_\mathrm{K}(\varpi).
\label{eqn:p0}
\end{equation}
The magnetic field strength $B_\mathrm{0}(\varpi)$ on the mid-plane is determined by using a ratio of the pressure and the magnetic pressure, $\beta\equiv p_\mathrm{0}(\varpi)/p_\mathrm{0,mag}(\varpi)$ where $p_\mathrm{0,mag}(\varpi) = B_\mathrm{0}^{2}(\varpi)/8\pi$,
\begin{equation}
B_\mathrm{0}(\varpi) = \sqrt{\frac{8\pi p_\mathrm{0}(\varpi)}{\beta}} = \sqrt{\frac{32\pi c}{3\alpha\beta}}\bar{\kappa}^{-1/2}_{\nu}(\varpi)\Omega^{1/2}_\mathrm{K}(\varpi)
\label{eqn:B0}
\end{equation}
where $\beta = 10$ is the assumption for our study.
Note that $B_\mathrm{0}(\varpi)$ is dominated by the toroidal magnetic field inside the disk.
For an inner region of optically thick neutrino driven accretion flow, the pressure from radiation and neutrino is dominated because of the high temperature.
It is therefore, for our analytical solution, the relation between the total energy density and temperature can be simplified to $\epsilon_\mathrm{0}(\varpi) = (11/4) a  T_\mathrm{0}^{4}(\varpi) + (7/8) a  T_\mathrm{0}^{4}(\varpi) = (29/8) a  T_\mathrm{0}^{4}(\varpi)$ where $a$ is the radiation constant \citep{2002ApJ...579..706D}.
Combining with Equation~\ref{eqn:epsilon0}, the temperature is expressed as
\begin{equation}
T_\mathrm{0}(\varpi) = \left(\frac{32c}{29\alpha a}\bar{\kappa}^{-1}_\nu(\varpi)\Omega_\mathrm{K}(\varpi)\right)^{1/4}.
\label{eqn:T}
\end{equation}
where the Rosseland mean opacity $\bar{\kappa}_{\nu}(\varpi)$ is determined by the total opacity due to neutrino-neutron and neutrino-proton scatterings which is formulated by \citet{2002ApJ...579..706D},
\begin{equation}
\bar{\kappa}_{\nu}(\varpi) = \kappa_{\nu0}\left(\frac{k_\mathrm{B}T_\mathrm{0}(\varpi)}{m_\mathrm{e}c^{2}}\right)^{2}
\label{eqn:opacity}
\end{equation}
where $\kappa_{\nu0} = 5.03\times 10^{-20}\,\mathrm{cm^{2} g^{-1}}$ for $k_\mathrm{B}T_\mathrm{0}(\varpi)\gg m_\mathrm{e} c^{2}$, $k_\mathrm{B}$ is the Boltzmann constant, and $m_\mathrm{e}$ is the electron mass \citep[see also][]{1964PhRv..136.1547B}.
Here the opacity depends on the temperature in contrast to the accretion disk model \citep{ebisuzaki+tasjima2019} in which the opacity is dominated by the Thomson scattering process.
Substituting Equation \ref{eqn:opacity} into Equation \ref{eqn:T}, the temperature is given by
\begin{align}
T_\mathrm{0}(\varpi) & = \left(\frac{32 m^{2}_\mathrm{e}c^{5}}{29\alpha a \kappa_\mathrm{\nu0} k_\mathrm{B}^{2}}\right)^{1/6}\Omega^{1/6}_\mathrm{K}(\varpi)\notag \\
 & = 3.61\times 10^{11} \left(\frac{\alpha}{0.1}\right)^{-1/6} \left(\frac{M}{\mathrm{M}_{\odot}}\right)^{-1/6} \left(\frac{\varpi}{\mathrm{r}_\mathrm{s}}\right)^{-1/4}\,\hbox{[K]}.
\label{eqn:Tdisk}
\end{align}
where $r_\mathrm{s}\equiv 2GM/c^{2}$ is the Schwarzschild radius.
It is therefore the opacity $\bar{\kappa}_{\nu}(\varpi)$ is expressed as
\begin{align}
\bar{\kappa}_{\nu}(\varpi) & = \left(\frac{32\kappa_{\nu0}^{2} k_\mathrm{B}^{4}}{29\alpha a m^{4}_\mathrm{e} c^{7}}\right)^{1/3}\Omega^{1/3}_\mathrm{K}(\varpi)\notag\\
 & = 1.86\times 10^{-16} \left(\frac{\alpha}{0.1}\right)^{-1/3} \left(\frac{M}{\mathrm{M}_{\odot}}\right)^{-1/3} \left(\frac{\varpi}{\mathrm{r}_\mathrm{s}}\right)^{-1/2}\,\hbox{[$\mathrm{cm}^{2}\,\mathrm{g}^{-1}$]}.
\label{eqn:kappa}
\end{align}
Substituting Equation \ref{eqn:kappa} into Equations \ref{eqn:scaleheight},\ref{eqn:ro0},\ref{eqn:epsilon0}, and \ref{eqn:B0}, we have
\begin{align}
H(\varpi) & =  \left(\frac{3\dot{M}}{16\pi c}\right)\left(\frac{32\kappa^{2}_{\nu0}k^{4}_\mathrm{B}}{29\alpha a m^{4}_\mathrm{e}c^{7}}\right)^{1/3}\Omega^{1/3}_\mathrm{K}(\varpi)\notag\\
 & = 7.36\times 10^{5} \left(\frac{\alpha}{0.1}\right)^{-1/3}\left(\frac{\dot{M}}{\dot{\mathrm{M}}_{\odot}}\right)\left(\frac{M}{\mathrm{M}_{\odot}}\right)^{-1/3}\left(\frac{\varpi}{\mathrm{r}_\mathrm{s}}\right)^{-1/2}\,\hbox{[$\mathrm{cm}$]}.
\label{eqn:H}
\end{align}
\begin{align}
\rho_\mathrm{0}(\varpi) & = \left(\frac{928\pi^{2} a m^{4}_\mathrm{e} c^{10}}{27\dot{M}^{2}\kappa^{2}_\mathrm{\nu0}k^{4}_\mathrm{B}}\right)\Omega^{-2}_\mathrm{K}(\varpi)\notag\\
 & = 5.52\times 10^{10}\left(\frac{\dot{M}}{\dot{\mathrm{M}}_{\odot}}\right)^{-2}\left(\frac{M}{\mathrm{M}_{\odot}}\right)^{2}\left(\frac{\varpi}{\mathrm{r} _\mathrm{s}}\right)^{3}\,\hbox{[$\mathrm{g}\,\mathrm{cm}^{-3}$]}.
\label{eqn:ro}
\end{align}
\begin{align}
\epsilon_\mathrm{0}(\varpi) & = \left(\frac{58 a m^{4}_\mathrm{e} c^{10}}{\alpha^{2}\kappa^{2}_{\nu0}k^{4}_\mathrm{B}}\right)^{1/3}\Omega^{2/3}_\mathrm{K}(\varpi)\notag\\
 & = 4.63\times 10^{32}\left(\frac{\alpha}{0.1}\right)^{-2/3}\left(\frac{M}{\mathrm{M}_{\odot}}\right)^{-2/3}\left(\frac{\varpi}{\mathrm{r}_\mathrm{s}}\right)^{-1}\,\hbox{[$\mathrm{erg}\,\mathrm{cm}^{-3}$]}.
\end{align}
\begin{align}
B_\mathrm{0}(\varpi) & = \left(\frac{8\pi}{3\beta}\right)^{1/2}\left(\frac{58 a m^{4}_\mathrm{e} c^{10}}{\alpha^{2}\kappa^{2}_{\nu0}k^{4}_\mathrm{B}}\right)^{1/6}\Omega^{1/3}_\mathrm{K}(\varpi)\notag\\
 & = 1.97\times 10^{16}\left(\frac{\beta}{10}\right)^{-1/2}\left(\frac{\alpha}{0.1}\right)^{-1/3}\left(\frac{M}{\mathrm{M}_{\odot}}\right)^{-1/3}\left(\frac{\varpi}{\mathrm{r}_\mathrm{s}}\right)^{-1/2}\,\hbox{[$\mathrm{G}$]}.
\label{eqn:B}
\end{align}
where $\dot{\mathrm{M}}_\odot\equiv \mathrm{M}_\odot / \mathrm{s}$.
Finally, the radial inflow velocity $- v_\varpi(\varpi) = \dot{M} / 2\pi \varpi\Sigma(\varpi) = \dot{M} / 4\pi \varpi\rho_\mathrm{0}(\varpi) H(\varpi)$ becomes
\begin{align}
- v_{\varpi}(\varpi) & = \left(\frac{9\dot{M}^{2}\alpha}{256\pi^{2} c^{2}\varpi}\right)\left(\frac{32\kappa^{2}_{\nu0}k^{4}_\mathrm{B}}{29 \alpha a m^{4}_\mathrm{e} c^{7}}\right)^{2/3}\Omega^{5/3}_\mathrm{K}(\varpi)\notag\\
  & = 1.31\times 10^{10} \left(\frac{\alpha}{0.1}\right)^{1/3} \left(\frac{\dot{M}}{\dot{\mathrm{M}}_{\odot}}\right)^{2} \left(\frac{M}{\mathrm{M}_{\odot}}\right)^{-5/3}\left(\frac{\varpi}{\mathrm{r}_\mathrm{s}}\right)^{-7/2}\,\hbox{[$\mathrm{cm}\,s^{-1}$]}.
\end{align}
%
The radial profile of the NDAF disk properties is shown in Figure\,\ref{fig:diskmodels}.
Note that our analytical solution is valid in the close proximity of BHs where the radiation pressure is dominated, which turns out to be unstable against both a secular instability ($\partial\dot{M}/\partial\Sigma < 0$) and a thermal instability ($\partial T_\mathrm{0}/\partial \Sigma = 0$) \citep{1998bhad.conf.....K,2007ApJ...664.1011J}.
However, it may not be subject to those instabilities if a toroidal magnetic field could hold the radiation-pressure-dominated disk \citep{2003A&A...407..403P,2007MNRAS.375.1070B,Zheng:2011kl}.
The role of such strong toroidal magnetic fields in the NDAF disks is an issue awaiting further investigation.
%

\subsection{Burst emissions of the electro-magnetic pulses}
We consider that the wavelength of the emitted electro-magnetic (EM) pulses is of the order of the size of the density fluctuations generated in the disk.
When the generation of the density fluctuations is regulated by the distance between two Alfv\'en singularities for the unstable non-axisymmetric mode in magneto-rotational instability \citep[MRI:][]{1995ApJ...445..767M}, the wavelength of EM wave pulses becomes
\begin{equation}
\lambda(\varpi) = \frac{\cal C_\mathrm{A}(\varpi)}{\cal C_\mathrm{s}(\varpi)}\frac{\Omega_\mathrm{K}(\varpi)}{\cal R(\varpi)}H(\varpi)
\label{eqn:lambda_def}
\end{equation}
where ${\cal C_\mathrm{A}(\varpi)}=B_\mathrm{0}(\varpi)/\sqrt{4\pi\rho_\mathrm{0}(\varpi)}$ is the Alfv\'en speed on the mid-plane in the disk and the shear rate ${\cal R}(\varpi) = -(\varpi/2)\left[d\Omega_\mathrm{K}(\varpi)/d\varpi\right]$ becomes
\begin{equation}
{\cal R(\varpi)} = \frac{3}{4}\Omega_\mathrm{K}(\varpi).
\label{eqn:A}
\end{equation}
By using Equations \ref{eqn:ro} and \ref{eqn:B}, the Alfv\'en speed ${\cal C_\mathrm{A}(\varpi)}$ becomes
\begin{align}
{\cal C_\mathrm{A}(\varpi)} & = 2.36\times 10^{10} \left(\frac{\beta}{10}\right)^{-1/2}\left(\frac{\alpha}{0.1}\right)^{-1/3} \left(\frac{\dot{M}}{\dot{\mathrm{M}}_{\odot}}\right) \left(\frac{M}{\mathrm{M}_{\odot}}\right)^{-4/3}\left(\frac{\varpi}{\mathrm{r}_\mathrm{s}}\right)^{-2}\,\hbox{[$\mathrm{cm}\,s^{-1}$]}.
\label{eqn:ca}
\end{align}
Using ${\cal C}_\mathrm{s}(\varpi) = H(\varpi)\Omega_\mathrm{K}(\varpi)$ and substituting Equations \ref{eqn:H}, \ref{eqn:A}, and \ref{eqn:ca} into Equation \ref{eqn:lambda_def},
\begin{align}
\lambda(\varpi) & = \frac{4}{3} {\cal C}_\mathrm{A}(\varpi) \Omega^{-1}_\mathrm{K}(\varpi)\notag\\
 & = 4.39\times 10^{5} \left(\frac{\beta}{10}\right)^{-1/2}\left(\frac{\alpha}{0.1}\right)^{-1/3} \left(\frac{\dot{\mathrm{M}}}{\dot{\mathrm{M}}_{\odot}}\right) \left(\frac{M}{
 \mathrm{M}_{\odot}}\right)^{-1/3}\left(\frac{\varpi}{\mathrm{r}_\mathrm{s}}\right)^{-1/2}\,\hbox{[$\mathrm{cm}$]}.
\label{eqn:lambda}
\end{align}
%

%
The timescale of emergence of an EM wave pulse on the disk $\tau_\mathrm{wave}$ is given by the half-thickness of the disk divided by the local Alfv\'en speed.
By using Equations \ref{eqn:ca} and \ref{eqn:H},
\begin{align}
\tau_\mathrm{wave} & =  \frac{H(\varpi)}{\cal C_\mathrm{A}(\varpi)} = \sqrt{\frac{\beta}{2}}\,\Omega^{-1}_\mathrm{K}(\varpi)\notag\\
 & = 3.11\times 10^{-5} \left(\frac{\beta}{10}\right)^{1/2} \left(\frac{\dot{M}}{\dot{\mathrm{M}}_{\odot}}\right)^{-1} \left(\frac{M}{\mathrm{M}_{\odot}}\right) \left(\frac{\varpi}{\mathrm{r}_\mathrm{s}}\right)^{3/2}\,\hbox{[$\mathrm{s}$]}.
\label{eqn:tau_wave}
\end{align}
%

The power of an EM wave pulse ${\cal P}_\mathrm{wave}(\varpi)$ propagating along the jet is estimated by the magnetic energy emerging from the disk in the vertical direction divided by $\tau_\mathrm{wave}$, and therefore
\begin{equation}
{\cal P}_\mathrm{wave}(\varpi)=\frac{B^{2}_\mathrm{0}(\varpi)}{8\pi} \frac{H(\varpi) d{\cal A}(\varpi)}{\tau_\mathrm{wave}}
\label{eqn:Pwave}
\end{equation}
where $d{\cal A}(\varpi)$ is the unit surface area on the disk.
Substituting Equations~\ref{eqn:H}, \ref{eqn:B}, and \ref{eqn:ca} into ${\cal P}_\mathrm{wave}(\varpi)$ dividing by $d{\cal A}(\varpi)$, the energy flux of EM wave pulses ${\cal F}_\mathrm{wave}(\varpi)$ is expressed as
\begin{align}
{\cal F}_\mathrm{wave}(\varpi) & = \frac{{\cal P}_\mathrm{wave}(\varpi)}{d{\cal A}(\varpi)} = \frac{{\cal C_\mathrm{A}(\varpi)} B^{2}_\mathrm{0}(\varpi)}{8\pi} = \frac{\dot{M}}{\pi\alpha\sqrt{8\beta^{3}}}\Omega^{2}_\mathrm{K}(\varpi) \notag\\
 &  = 3.65\times 10^{41} \left(\frac{\beta}{10}\right)^{-3/2} \left(\frac{\alpha}{0.1}\right)^{-1} \left(\frac{\dot{M}}{\dot{\mathrm{M}}_{\odot}}\right) \left(\frac{M}{\mathrm{M}_{\odot}}\right)^{-2} \left(\frac{\varpi}{\mathrm{r}_\mathrm{s}}\right)^{-3}\,\hbox{[$\mathrm{erg}\,\mathrm{cm}^{-2}\,\mathrm{s}^{-1}$]}.
 \label{eqn:Fwave}
\end{align}
The total wave luminosity from the entire disk is calculated by integrating ${\cal F}_\mathrm{wave} d{\cal A}(\varpi)$ over the radius:
\begin{align}
L_\mathrm{wave} & = \int_{\varpi_\mathrm{in}}^{\infty} 2 {\cal F}_\mathrm{wave}(\varpi) 2\pi \varpi d\varpi = \frac{\dot{M}}{\alpha}\left(\frac{2}{\beta^{3}}\right)^{1/2}\left(\frac{GM}{\varpi_\mathrm{in}}\right) = \left(\frac{1}{18\alpha^{2}\beta^{3}}\right)^{1/2}\dot{M}c^{2}\notag\\
 & = 1.33\times 10^{53} \left(\frac{\beta}{10}\right)^{-3/2} \left(\frac{\alpha}{0.1}\right)^{-1} \left(\frac{\dot{M}}{\dot{\mathrm{M}}_{\odot}}\right)\,\hbox{[$\mathrm{erg}\,\mathrm{s}^{-1}$]}.
\label{eqn:Lw}
\end{align}
Here we assume that the inner edge of the disk is truncated at the innermost stable circular orbit  (ISCO)\footnote{Although we use the Newtonian gravitational potential, it is convenient to set it as the inner boundary of the disk around a non-rotating BH.}, thus $\varpi_\mathrm{in} = 3 \mathrm{r}_\mathrm{s}$.
Note that a factor of $2$ comes from both upper and lower sides of the disk.
%

Likewise, from Equation \ref{eqn:radiation_flux}, the neutrino energy flux  ${\cal F}_\nu$ becomes
\begin{align}
{\cal F}_{\nu}(\varpi) & = 1.22\times 10^{42} \left(\frac{\dot{M}}{\dot{\mathrm{M}}_{\odot}}\right) \left(\frac{M}{\mathrm{M}_{\odot}}\right) \left(\frac{\varpi}{\mathrm{r}_\mathrm{s}}\right)^{-3}\,\hbox{[$\mathrm{erg}\,\mathrm{cm}^{-2}\,\mathrm{s}^{-1}$]}, 
\label{eqn:Frad}
\end{align}
and the neutrino luminosity $L_{\nu}$ is calculated by integrating $F_\nu$ over the entire disk:
\begin{align}
L_\nu & = \int_{\varpi_\mathrm{in}}^{\infty} 2 {\cal F}_\nu(\varpi) 2\pi \varpi d\varpi = \frac{3\dot{M}}{2}\frac{GM}{\varpi_\mathrm{in}} = \frac{1}{4}\dot{M}c^{2}\notag\\
 & = 4.47\times 10^{53} \left(\frac{\dot{M}}{\dot{\mathrm{M}}_{\odot}}\right)\,\hbox{[$\mathrm{erg}\,\mathrm{s}^{-1}$]}.
\label{eqn:Lnu}
\end{align}
The ratio of the wave luminosity to the neutrino luminosity $L_\mathrm{wave}/L_\nu$ becomes
\begin{equation}
\frac{L_\mathrm{wave}}{L_{\nu}} = \left(\frac{8}{9\alpha^{2}\beta^{3}}\right)^{1/2}= 0.298 \left(\frac{\alpha}{0.1}\right)^{-1}\left(\frac{\beta}{10}\right)^{-3/2}
\end{equation}
which is consistent with the value reported as $\mathcal{O}(1)$ in \citet{ebisuzaki+tasjima2019}, suggesting that the wave luminosity can be the primary energy source of GRBs, SNe, and HNe.
%

\subsection{Propagation of the electro-magnetic wave pulses along the jet}
%
The magnetic fields contained within the progenitor are twisted and amplified via the MRI or dynamo action in the NDAF disk.
Such a strong toroidal magnetic field component is eventually emerged from the surface of the disk via the magnetic buoyancy and converted into a large-scale vertical magnetic field, which is known as a magnetic tower \citep{1996MNRAS.279..389L,2004ApJ...605..307K,2006ApJ...647.1192U}.
In addition, such explosive magnetic flux ejections from the magnetized accretion disks are expected to occur repeatedly \citep{1990ApJ...350..295S}.
Once large-amplitude Alfv\'enic wave pulses (which turn into EM wave pulses) have launched from the NDAF, it propagates along the vertical magnetic field confined within a funnel shape in the jet as a wave packet at nearly the speed of light (See Figure~\ref{fig:schematic}).
Assuming that EM wave pulses do not interact with the surroundings outside the jet, the energy injected by the electric field $E_\mathrm{0}(\varpi)$ within each propagating pulse is expressed as:
\begin{equation}
E_\mathrm{0}(\varpi) = \sqrt{\frac{4\pi{\cal F}_\mathrm{wave}(\varpi)}{c}}.
\end{equation}
From Equation \ref{eqn:Fwave}, it becomes
\begin{equation}
E_\mathrm{0}(\varpi) = 1.24\times  10^{16}\left(\frac{\beta}{10}\right)^{-3/4} \left(\frac{\alpha}{0.1}\right)^{-1/2} \left(\frac{\dot{M}}{\dot{\mathrm{M}}_{\odot}}\right)^{1/2} \left(\frac{M}{\mathrm{M}_{\odot}}\right)^{-1} \left(\frac{\varpi}{\mathrm{r}_\mathrm{s}}\right)^{-3/2}\,\hbox{[$\mathrm{dyn\,esu^{-1}}$]}.
\label{eqn:Ew}
\end{equation}
By analogy with an important parameter in the intense laser-plasma interactions, we employ the wakefield strength parameter \citep{1988ApPhL..53.2146S}:
\begin{equation}
a_\mathrm{0}(\varpi) = e A_\mathrm{0}/m_\mathrm{e}c^{2}
\end{equation}
where $A_\mathrm{0}\equiv c E_\mathrm{0}(\varpi)/\omega$ is the amplitude of the vector potential at the base of the jet and $\omega = 2\pi c/\lambda(\varpi)$ is the angular frequency of the EM wave pulse in the jet.
Note that we assume the propagation speed of the EM wave pulse to be the speed of light.
This assumption holds the most of the cases including jets from neutrino driven accretion flows.
The parameter $a_\mathrm{0}(\varpi)$ is basically the normalized vector potential amplitude of the EM wave pulse at the base of the jet.
By substituting Equation \ref{eqn:Ew}, 
\begin{equation}
a_\mathrm{0} (\varpi) = 5.07\times 10^{17} \left(\frac{\beta}{10}\right)^{-5/4} \left(\frac{\alpha}{0.1}\right)^{-4/3} \left(\frac{\dot{M}}{\dot{\mathrm{M}}_{\odot}}\right)^{3/2} \left(\frac{M}{\mathrm{M}_{\odot}}\right)^{-4/3} \left(\frac{\varpi}{\mathrm{r}_\mathrm{s}}\right)^{-2}.
\label{eqn:azero}
\end{equation}
%

%
The physical parameters within the jet as a function of the distance from the disk $z$ determines how the electro-magnetic (EM) wave pulses propagate through the jet.
However, the structure of jet emanated from the NDAF disk and penetrating though a progenitor ambient medium is not easy to characterize.
Therefore, we assume that the so-called jet collimation profile is determined by the normalized distance $z/\varpi_\mathrm{0}$ in powers of $\phi$ for simplicity, and therefore the radius of the jet can be expressed as
\begin{equation}
R(\varpi_\mathrm{0},z)=  \varpi_\mathrm{0} \left[1 + \left(z / \varpi_\mathrm{0}\right)^{\phi}\right]
\end{equation}
where $\varpi_\mathrm{0}$ is the radius at the base of the jet.
Here we choose $\phi=1/2$ and $\phi=1$, namely, a parabolic shape for a collimated jet and a conical shape for an uncollimated wind (its opening angle of $45$ degrees), respectively.
Assuming the magnetic flux within the area ${\cal A}(z) = \pi R^{2}(\varpi_\mathrm{0},z)$ is conserved along the jet, $B_\mathrm{0}(\varpi_\mathrm{in}){\cal A}(0) = B(z) {\cal A}(z)=\mathrm{const.}$, it is therefore the magnetic field strength along the jet $B(z)$ can be expressed as
\begin{align}
B(z) & = B_\mathrm{0}{\cal A}(0)/{\cal A}(z).
\label{eqn:Bz}
\end{align}

The plasma number density $n_\mathrm{p}(z)$ in the jet is estimated by assuming the ratio between the total kinetic luminosity and the total neutrino luminosity is constant along the jet,
\begin{equation}
L_\mathrm{kinetic}=n_\mathrm{p}(z)\mu m_\mathrm{p} c^{3}\Gamma^{2}{\cal A}(z) = \xi L_{\nu}
\label{eqn:L_kinetic}
\end{equation}
where the mean molecular weight $\mu = 2.34$, and we set the constant jet bulk Lorentz factor of $\Gamma=400$ \citep[which is slightly higher than the averaged value of][]{2018A&A...609A.112G} and we choose $\xi=0.1$.
From the charge neutrality in the jet, the electron number density $n_\mathrm{e}(z)$ can be expressed as
\begin{equation}
n_\mathrm{e}(z) = n_\mathrm{p}(z) = \frac{\xi L_{\nu}}{\mu m_\mathrm{p} c^{3}\Gamma^{2}{\cal A}(z)}.
\end{equation}
%

\begin{figure}
\epsscale{0.85}
\plotone{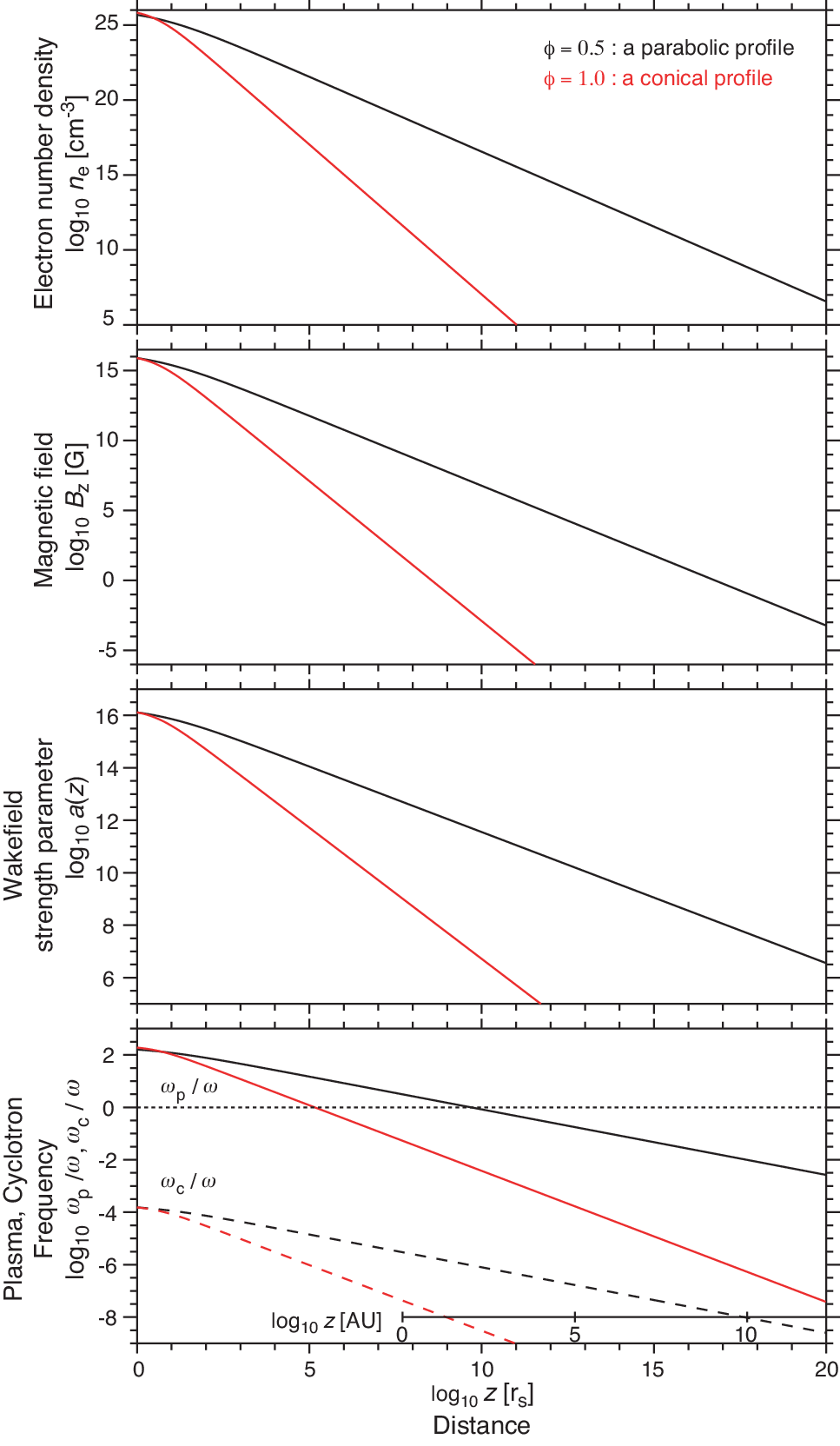}
\caption{The vertical structure of the jet properties for $\phi=0.5$ (black) and $\phi=1$ (red).  The electron number density, the vertical magnetic field strength, the wakefield strength parameter ($\approx$ the Lorentz factor of electrons), and the angular frequency of electro-magnetic (EM) wave pulses of the jet ($\omega$) in comparison with both the cyclotron frequency ($\omega_{\mathrm{c}}$) and the plasma frequency ($\omega_{\mathrm{p}}$) are shown from top to bottom panels, respectively.  The mass of a black hole is $M = 3 \mathrm{M}_\odot$ and the ratio between the total kinetic luminosity of the jet ($L_\mathrm{kinetic}$) and the total neutrino luminosity ($L_\mathrm{\nu}$) is assumed to be $\xi\equiv L_\mathrm{kinetic}/L_\mathrm{\nu}=0.1$.  The region of $\omega_\mathrm{p}/\omega > 1$ is the evanescent region but the presence of magnetic field makes some of the magnetic waves not evanescent, whereas the region of $\omega_\mathrm{p}/\omega < 1$ is the propagation region for EM wave pulses.}
\label{fig:jet}
\end{figure}

From Equation~\ref{eqn:azero}, the wakefield strength parameter in the jet $a(z)$ becomes much greater than the unity, $a(z)\gg 1$, trapped electrons in the EM wave pulses become ultra-relativistic and therefore the Lorentz factor of electrons becomes $\gamma_\mathrm{e}(z) \approx a(z)$ \citep{2017NCimR..40...33T} where $\gamma_\mathrm{e}(z) = 1/\sqrt{1- (|\vec{v}_\mathrm{e}| / c)^{2}}$ is the Lorentz factor of trapped electrons with the velocity of electrons $\vec{v}_\mathrm{e}$ in the wake.
By assuming the base of the jet is located at $\varpi_\mathrm{0} = \varpi_\mathrm{in}$ and the energy flux of EM wave pulses is conserved along the jet, ${\cal A}(z) a^{2}(z) = {\cal A}(0) a^{2}_\mathrm{0}(\varpi_\mathrm{0}) = \mathrm{const.}$, the wakefield strength parameter can be expressed as
\begin{equation}
\gamma_\mathrm{e}(z) \approx a(z) = a_\mathrm{0}(\varpi_\mathrm{0}) \sqrt{{\cal A}(0)/{\cal A}(z)}.
\end{equation}
Since $\vec{v}_\mathrm{e}$ is mostly perpendicular to the bulk velocity of the jet, the plasma frequency is expressed as
\begin{equation}
\omega_\mathrm{p}(z) = \sqrt{4\pi n_\mathrm{e}(z) e^{2}/m_\mathrm{e}\gamma(z)}
\label{eqn:omegap}
\end{equation}
where $\gamma(z)=\gamma_\mathrm{e}(z)\Gamma$ is the Lorentz factor of the combined velocities in the jet.
Note that the cyclotron frequency in the jet is $\omega_\mathrm{c}(z)=\sqrt{e B(z)/m_\mathrm{e}c\gamma(z)}$.
%
Finally, the vertical structure of the jet properties such as the electron number density, the vertical magnetic field strength, the Lorentz factor of electrons, and the EM wave frequency of the jet in comparison with both the plasma frequency and the cyclotron frequency are shown in Figure\,\ref{fig:jet}.
%

\section{Neutrino spectrum of NDAF disk} \label{sec:results1}
\begin{figure}
\epsscale{0.85}
\plotone{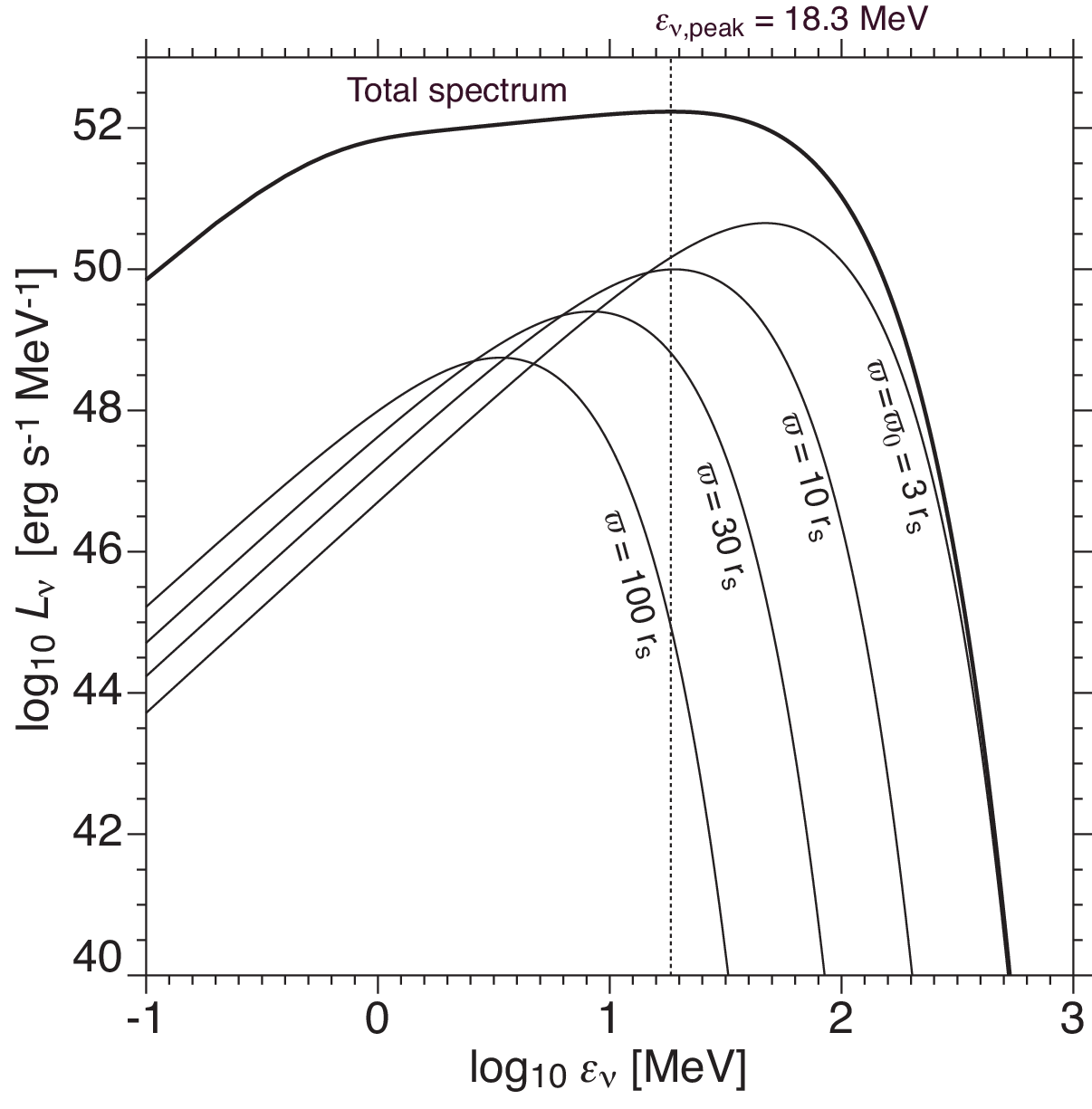}
\caption{Neutrino spectrum of NDAF of $L_{\nu}$ with the mass accretion rate of $\dot{M}=1.0\,\mathrm{M}_\odot/\mathrm{s}$.  The peak value of neutrino energy $\epsilon_{\nu\mathrm{, peak}}$ is $18.3~\mathrm{MeV}$, which is consistent with the previously estimated values \citep{2016PhRvD..93l3004L}.  This peak emission is originated from the inner region of NDAF disks ($\varpi\lsim 10~r_\mathrm{s}$).}
\label{fig:neutrino_spectrum}
\end{figure}

Assuming the disk is a blackbody source for neutrinos at each radius, the neutrino energy flux per unit energy interval per solid angle to be 
\begin{equation}
{\cal B}_{\nu}(\varepsilon_{\nu},T_\nu(\varpi)) = \frac{4\varepsilon^{3}_{\nu}/h^{3}c^{2}}{\exp{\left[\left(\varepsilon_{\nu} - \mu_\nu\right)/k_\mathrm{B}T_\nu(\varpi)\right]} + 1}.
\end{equation}
where $\varepsilon_{\nu}$ is the energy of neutrino, $T_{\nu}(\varpi)$ is the effective temperature of neutrino at each radius, and $\mu_\nu$ is the chemical potential of neutrinos.
We derive $T_{\nu}(\varpi)$ by using the relation ${\cal F}_{\nu}(\varpi) = (7/8)a T^{4}_{\nu}(\varpi)$.
Assuming that the chemical potential of neutrinos can be ignored, $\mu_{\nu}=0$ \citep{2007ApJ...662.1156K}, we compute the emergent luminosity of neutrino  $L_\nu(\varepsilon_{\nu}) = 4\pi^{2}{\cal B}_\nu(\varepsilon_{\nu},T_\nu(\varpi))\varpi d\varpi$ as a function of the energy of neutrino and plot the spectra as shown in Figure \ref{fig:neutrino_spectrum}.
The spectra extend more than $100~\mathrm{MeV}$ and the peak value of neutrino energy is $18.3~\mathrm{MeV}$, which is consistent with the previous estimation of NDAF disks \citep{2016PhRvD..93l3004L}.
This peak emission, which is higher than a peak neutrino energy expected from the standard supernova with a hot neutron star \citep{1987ApJ...318..288M}, is originated from the inner region of NDAF disks ($\varpi\lsim 10~r_\mathrm{s}$).
Since the higher neutrino energy becomes the larger neutrino cross-sections in water becomes \citep[e.g.,][]{2003PhLB..564...42S}, water Cherenkov detectors, like Super-Kamiokande (SK), could have more chance to detect neutrinos from NDAFs.

\section{EM Wave Pulse in Jet} \label{sec:results2}
\subsection{Explosion energy by waves and accreted mass onto a black hole}
\begin{figure}
\epsscale{0.85}
\plotone{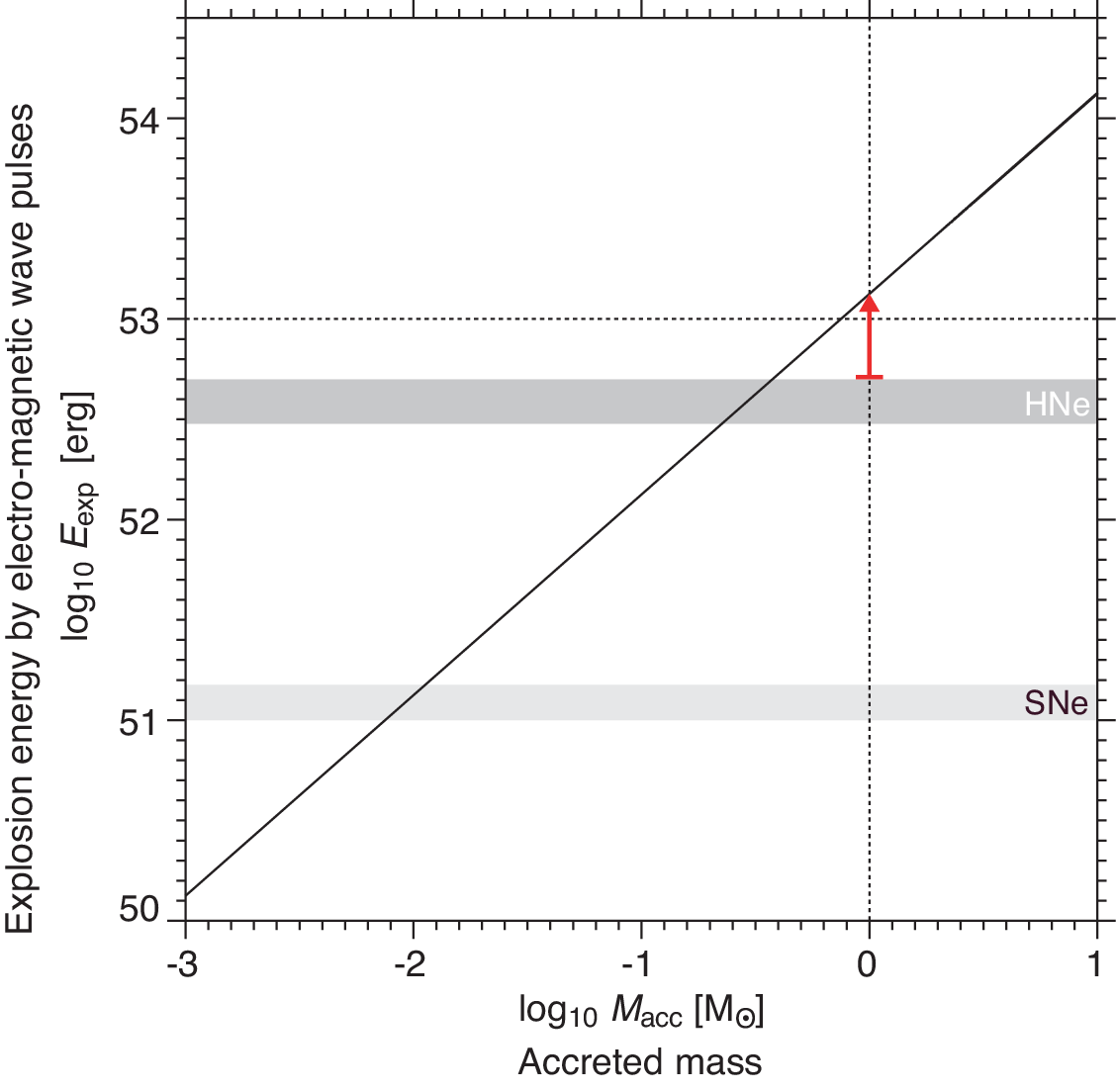}
\caption{Explosion energy emitted by EM wave pulses from NDAF and the accreted mass onto a black hole ($\alpha = 0.1$ and $\beta = 10$).  Two shaded bands represent typical explosion energy for supernovae (SNe) and hypernovea (HNe) \citep{2005hedl.book...81N,2013ARA&A..51..457N}.  If the accreted mass becomes one solar mass, the explosion energy could exceed $10^{53}\,\mathrm{erg}$ which is strong enough for reversing collapsing cores of massive stars into explosion and therefore our model is relevant for HNe as indicated by a red arrow.}
\label{fig:wave_power}
\end{figure}

We have estimated the explosion energy released as EM wave pulses $E_\mathrm{exp}$ by integrating Equation \ref{eqn:Lw} over time during the growth of a BH
\begin{equation}
E_\mathrm{exp} = \int_{0}^{\tau} L_\mathrm{wave} dt = \frac{c^{2}}{6\alpha}\left(\frac{2}{\beta^{3}}\right)^{1/2}\int_{0}^{\tau}\dot{M} dt = \frac{c^{2}}{6\alpha}\left(\frac{2}{\beta^{3}}\right)^{1/2} M_\mathrm{acc}
\end{equation}
where $M_\mathrm{acc}$ is the accreted mass.
Figure~\ref{fig:wave_power} shows the relation between the explosion energy by EM wave pulses $E_\mathrm{exp}$ and the accumulated mass $M_\mathrm{acc}$.
If the mass of progenitors would be at most several tens of the solar mass, the very high accretion rate phase ($\dot{M} \sim 1 \mathrm{M}_{\odot} \rm s^{-1}$) in collapsars will last only several seconds to tens of seconds.
It is important to note that the explosion energy reaches high as $10^{53}~\mathrm{erg}$ if the accreted mass becomes $1\,\mathrm{M}_\odot$, and therefore our model is relevant for HNe.

\subsection{Acceleration of high energy particles in the jet}
\begin{figure}
\epsscale{0.85}
\plotone{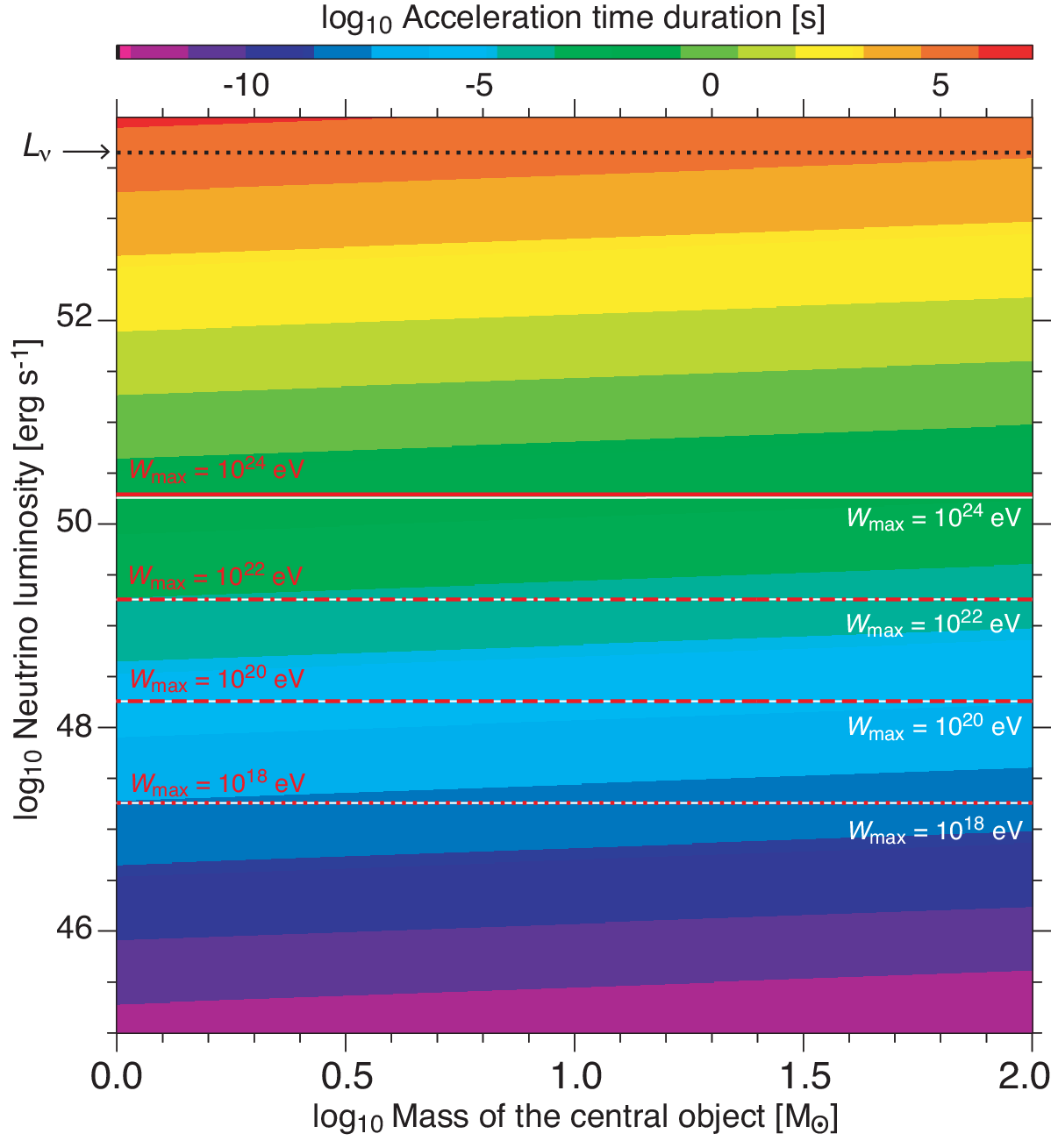}
\caption{Maximum energy of accelerated protons ${\cal W}_\mathrm{max} = 10^{18}$, $10^{20}$, $10^{22}$, and $10^{24}~\mathrm{eV}$ for $\phi=1/2$ (white) and $\phi=1$ (red) in the jets from the NDAFs in relation to the mass of the central objects and neutrino luminosity.  Color-scale indicates the acceleration time duration defined as $\Delta z_\mathrm{w}/c$ which is independent to $\phi$.  A black dotted line shows the neutrino luminosity of NDAFs with the mass accretion rate of $\dot{M}=1.0\,\mathrm{M}_\odot/\mathrm{s}$, indicating that the energy of protons gained by the wakefield acceleration is sufficient for a source of the extremely high energy cosmic rays (EHECRs) and the super-EHECRs of $10^{22 - 23}~\mathrm{eV}$ \citep{Takahashi2000}.  The acceleration time required for the maximum energy of protons is not more than $\approx 10^{5}~\mathrm{s}$ and therefore the wakefield acceleration is efficient for generating the super-EHECRs as well.}
\label{fig:max_energy}
\end{figure}

The pondermotive force exerted on electrons by EM wave pulses generates longitudinal polarization of electron distribution which creates wakefields as illustrated in Figure\,\ref{fig:schematic}.
Charged particles are accelerated by the wakefield $E_\mathrm{TD} = m_\mathrm{e} \omega_\mathrm{p}(z) c / e$ \citep[the so-called Tajima-Dawson field:][]{1979PhRvL..43..267T}, and therefore the wakefield force $F_\mathrm{w}$ exerted on a charged particle in the non-relativistic regime is expressed as $F_\mathrm{w} = {\cal Z} e E_\mathrm{TD} = {\cal Z} m_\mathrm{e} \omega_\mathrm{p}(z) c$ where ${\cal Z}$ is the charge number.
In the relativistic regime, namely under the intense wakefield $a(z)\approx\gamma(z)\gg 1$ and high bulk Lorentz factor $\Gamma \ge 1$, the wakefield force  $F_\mathrm{w}$ becomes \citep{2017NCimR..40...33T}
\begin{equation}
F_\mathrm{w} = {\cal Z}\Gamma m_\mathrm{e} a(z) \omega_\mathrm{p}(z) c.
\label{eqn:F_w}
\end{equation}
Note that the accelerating force is the same for electrons (positrons) and protons.
The maximum energy ${\cal W}_\mathrm{max}$ gained by the accelerated charged particles is determined by integrating the work done by $F_\mathrm{w}$ over the acceleration distance {\bf $\Delta z_\mathrm{w}$}
\begin{equation}
{\cal W}_\mathrm{max} = \int_{z_\mathrm{w}}^{z_\mathrm{w}+\Delta z_\mathrm{w}} F_\mathrm{w} dz = {\cal Z} \Gamma m_\mathrm{e} c \int_{z_\mathrm{w}}^{z_\mathrm{w}+\Delta z_\mathrm{w}} \omega_\mathrm{p}(z) a(z) dz.
\label{eqn:wmax}
\end{equation}
Here we assume that $z_\mathrm{w}$ is the acceleration site on which the plasma becomes the underdense condition ($\omega > \omega_\mathrm{p}$) from the overdense condition ($\omega < \omega_\mathrm{p}$).
As shown in Figure \ref{fig:jet}, the acceleration site $z_\mathrm{w}$ depends on the vertical structure of the jet, more precisely the density stratification in the jet.
The energy gained for an electron (and a position) is most likely limited by the radiation loss such as the synchrotron and Bremsstrahlung emissions because both magnetic field and charged particles in the jet bend their trajectory.
Thus, after the onset of the wakefield acceleration at $z \gsim z_\mathrm{w}$, the synchrotron emissions by an accelerated electron (positron) result in an observational signatures of gamma-ray emissions.
On the other hand, the energy gained by a proton is determined by the coherent accelerating length available over the jet environment.
In the laser wakefield accelerations for $a_\mathrm{0} \gg 1$, the maximum coherent accelerating length is determined by a half of the pump depletion length $l_\mathrm{pd} = 2\sqrt{2} c \left(\omega^{2} / \omega^{3}_\mathrm{p}\right) a_\mathrm{0}$  where $\omega$ and $\omega_\mathrm{p}$ is the frequency of the laser beam and the plasma frequency, respectively \citep{Esarey:2009ks}.
This is an ideal condition for the maximum acceleration length, and therefore we remind the readers that the effective acceleration length could be shorter in reality.
In our study, we consider the maximum coherent accelerating length as the acceleration distance in the jets, namely $\Delta z_\mathrm{w} = l_\mathrm{pd}/2 =  \sqrt{2} c \left[\omega_\mathrm{0}^{2} / \omega^{3}_\mathrm{p}(z_\mathrm{w})\right] a(z_\mathrm{w})$ where $\omega_\mathrm{0} = 2\pi c/\lambda(\varpi_\mathrm{0})$ is the frequency of EM wave pulses at the base of the jet.
Since the integral of $\omega_\mathrm{p}(z) a(z)$ in Equation~\ref{eqn:wmax} is complicated, we shall reduce $\omega_\mathrm{p}(z)$ and $a(z)$ to the following forms: for $z\gg \varpi_\mathrm{0}$, ${\cal R}(\varpi_\mathrm{0},z) \approx \varpi_\mathrm{0}\left(z/\varpi_\mathrm{0}\right)^{\phi}$ leads to $\omega_\mathrm{p}(z) \approx \omega_\mathrm{p}(0) \varpi_\mathrm{0}^{\phi/2} z^{-\phi/2}$ and $a(z) \approx a_\mathrm{0}(\varpi_\mathrm{0}) \varpi_\mathrm{0}^{\phi} z^{-\phi}$.
After some of algebra Equation\,\ref{eqn:wmax} becomes
\begin{align}
{\cal W}_\mathrm{max} & \approx {\cal Z} \Gamma m_\mathrm{e} \omega_\mathrm{p00} a_\mathrm{00}  c \dot{M}^{3/4} \Omega^{2/3}_\mathrm{K}(\varpi_\mathrm{0}) \varpi_\mathrm{0}^{3\phi/2 - 1} \int_{z_\mathrm{w}}^{z_\mathrm{w}+\Delta z_\mathrm{w}} z^{-3\phi/2} dz,\notag\\
 & = {\cal W}_\mathrm{0} \phi_\mathrm{0}^{-1} \left[\left(z_\mathrm{w} + \Delta z_\mathrm{w}\right)^{\phi_\mathrm{0}} - z_\mathrm{w}^{\phi_\mathrm{0}}\right] \dot{M}^{3/4} \Omega^{2/3}_\mathrm{K}(\varpi_\mathrm{0}) \varpi_\mathrm{0}^{-\phi_\mathrm{0}}
\label{eqn:wmax3}
\end{align}
where $\phi_\mathrm{0}=1-\frac{3}{2}\phi$, ${\cal W}_\mathrm{0} = {\cal Z}\Gamma m_\mathrm{e} \omega_\mathrm{p00} a_\mathrm{00} c$, and the acceleration site and the acceleration distance are
\begin{equation}
z_\mathrm{w} =\left[\left(\frac{\omega_\mathrm{p00}}{\omega_\mathrm{00}}\right)\left(\frac{6}{c^{2}}\right)^{\frac{\phi-1}{2}}\dot{M}^{1/4}\left(GM\right)^{\frac{3\phi-4}{6}}\right]^{2/\phi},
\label{eqn:zw}
\end{equation}
\begin{equation}
\Delta z_\mathrm{w} = \left(\frac{1}{3}\right)^{3/2} \left(\frac{c^{4}}{4}\right) \left(\frac{\omega_\mathrm{00}}{\omega^{2}_\mathrm{p00}} a_\mathrm{00}\right)\dot{M}^{2}\left(GM\right)^{-1/3},
\label{eqn:dzw}
\end{equation}
{\bf and}
\begin{equation}
a_\mathrm{00} = \frac{e \kappa^{2/3}_{\nu\mathrm{0}} k^{4/3}_\mathrm{B}}{2^{7/12} 29^{1/3} \pi^{2} m^{7/3}_\mathrm{e} c^{35/6} \alpha^{5/6} \beta^{5/4} a^{1/3}},
\end{equation}
\begin{equation}
\omega_\mathrm{p00} = \frac{2^{31/24} 29^{1/6} \pi \sqrt{\xi L_\nu e} \alpha^{5/12} \beta^{5/8} c^{17/12} a^{1/6} m^{2/3}_\mathrm{e}}{\sqrt{\mu m_\mathrm{p} \Gamma^{3}} \kappa^{1/3}_{\nu\mathrm{0}} k^{2/3}_\mathrm{B}},
\end{equation}
\begin{equation}
\omega_\mathrm{00} = \frac{2^{5/6} 29^{1/3} \pi^{2} \alpha^{1/3} \beta^{1/2} a^{1/3} m^{4/3}_\mathrm{e} c^{13/3}}{\kappa^{2/3}_{\nu\mathrm{0}} k^{4/3}_\mathrm{B}}.
\end{equation}
If the accelerated ion were carbon and oxygen nucleus, the gain by such ions is ${\cal Z}$-times the value of proton.
By substituting Equations \ref{eqn:lambda}, \ref{eqn:azero}, and $\varpi_\mathrm{0} = \varpi_\mathrm{in}$, Equation~\ref{eqn:wmax3} becomes
\begin{equation}
{\cal W}_\mathrm{max} = {\cal W}_\mathrm{0} \phi_\mathrm{0}^{-1} \left[\left(z_\mathrm{w} + \Delta z_\mathrm{w}\right)^{\phi_\mathrm{0}} - z_\mathrm{w}^{\phi_\mathrm{0}}\right] \left(\frac{c^2}{6}\right)^{\phi_\mathrm{0}+1} \dot{M}^{3/4} \left(GM\right)^{-(\phi_\mathrm{0}+2/3)}.
\label{eqn:wmax4}
\end{equation}
Note that the maximum energy gain ${\cal W}_\mathrm{max}$ becomes exactly independent of the mass of central objects $M$ if $\phi=1$.
On the other hand, if $\phi=1/2$, it is less dependent on $M$ because $\Delta z_\mathrm{w}$ is independent of $\phi$.
From Equation\,\ref{eqn:Lnu}, the mass accretion rate is expressed by $\dot{M} = 4 {\cal L}_\nu / c^{2}$ where ${\cal L}_\nu$ is the neutrino luminosity constrained by the future observations, and therefore, substituting it into Equations\,\ref{eqn:zw}, \ref{eqn:dzw}, and \ref{eqn:wmax4}, we plot the relation between observables such as neutrino luminosity ${\cal L}_{\nu}$ and BH mass ${\cal M}$ for a given ${\cal W}_\mathrm{max}$ in Figure~\ref{fig:max_energy}.
This shows that the neutrino luminosity necessary for the maximum energy gain for $< 10^{24}~\mathrm{eV}$ is well-below the neutrino luminosity from the NDAF disks of $L_\nu = 4.47\times 10^{53}\,\mathrm{erg}\,\mathrm{s}^{-1}$.
Therefore the energy of protons gained by the wakefield acceleration seems to be sufficient for a source of both the extremely high energy cosmic rays (EHECRs) and the super-EHECRs of $10^{22 - 23}~\mathrm{eV}$ \citep{Takahashi2000}.
In addition, the acceleration time required for the maximum energy of protons for $< 10^{24}~\mathrm{eV}$ is not more than $\approx 1~\mathrm{s}$, and therefore the wakefield acceleration seems to be efficient for generating the super-EHECRs as well.
%

\section{Summary and Discussion}\label{sec:summary_dicussion} \label{sec:summary_and_discussion}
In this paper, we have derived the properties of NDAF disks analytically and estimated the energy extracted via electro-magnetic (EM) wave pulses from the NDAF disks for the first time.
We have found that the energy emitted by EM wave pulses becomes more than $10^{53}~\mathrm{erg}$ in the form of Poynting flux if the accreted mass reached $\sim 1 \mathrm{M}_\odot$.
Such energy is sufficient for reversing collapsing cores of massive stars into explosion.
Since NDAF disks can only be formed in collapsars under the condition in which the specific angular momentum of progenitors becomes $j \gsim 1.5\times 10^{16}\,(M/\mathrm{M}_\odot)\,\mathrm{cm^{2}\,s^{-1}}$, NDAF disks can explain the bifurcation between HNe and faint SNe \citep{2013ARA&A..51..457N}.
Therefore the explosion driven by EM wave pulses from NDAF disks is the plausible model for HNe.
The wave energy penetrating through the collapsing cores after the explosion seems to be enough to generate high-energy particles up to the super-extremely high energy cosmic rays (super-EHECRs) of $10^{22-23}\,\mathrm{eV}$ as a result of the wakefield acceleration for ions in the jets.
Meanwhile, accelerated electrons and positrons can be the source of gamma-ray emissions as well as non-thermal emissions in GRBs \citep{Takahashi2000}.
The neutrino spectra of NDAF disks are extended up to $100\,\mathrm{MeV}$ and its peak value of neutrino energy is $18.3~\mathrm{MeV}$, which is an order of magnitude larger than the threshold energy of neutrinos in the previous study for evaluating the detection efficiency of the water Cherenkov detectors of Super-Kamiokande (SK) \citep{1998ApJ...496..216T,2003APh....18..551N} and its cross-section for inverse beta decay reaches $\gsim 10^{-41}~\mathrm{cm}^{2}$ \citep{2011JPhCS.309a2028S}.
Therefore, neutrino signals from NDAF disks could be a primary target for SK.
If the SK could have good timing capabilities equivalent to the Keplerian rotation period at the inner-edge of NDAF disks, $\delta t = 2\pi/\Omega_\mathrm{K}(\varpi_\mathrm{0}) \sim 4.0\times 10^{-4}\,\left(M/\mathrm{M}_\odot\right)\,\mathrm{s}$, the detection of neutrino intensity modulation as a result of deformations of NDAF disks would constrain the origin of neutrino signals as well as the properties of the central BHs.
%

\subsection{Dependency on the jet collimation profile}
The collimation profile of the jets from NDAFs is unknown.
But it can be extrapolated from the knowledge of jets from active galactic nuclei (AGNs) such as a powerful radio jet in M87.
The collimation profile in the M87 jet is known to be a parabolic-shape constrained within a few hundreds of the Schwartzschild radii in the vicinity of a supermassive black hole \citep{2017PASJ...69...71H}.
This is why we use a parabolic-shape in the first place.
But we also use a conical-shape because accretion-disk winds, which are an alternative way to drive outflows, have been observed in AGNs \citep{Tombesi:2015cx}.
By assuming the jet collimation profile is inherent between $\phi = 1/2$ and $\phi=1$, we can investigate a wider variety of the density stratification in the jet from NDAFs.
The onset of the acceleration process between $\phi=1/2$ and $\phi=1$ is quite different, because we assume that the acceleration starts at the same density, magnetic field strength, and wakefield strength parameter according to the condition of $\omega_\mathrm{p}(z_\mathrm{w})/\omega_\mathrm{0}=1$, but at the different location depending on the jet collimation profile as shown in Figure~\ref{fig:jet}.
In our model, the acceleration site for $\phi=1$ is $z_{\mathrm{w},\phi=1} = 1.5\times 10^{6}~\mathrm{r_{s}}$ whereas that of $\phi=1/2$ is $z_{\mathrm{w},\phi=1/2} = 7.2\times 10^{11}~\mathrm{r_{s}}$.
As a result, the generation of high energy charged particles by the wakefield acceleration must be delayed due to the propagation time of $\delta t_{\phi=1} = z_{\mathrm{w},\phi=1}/c \approx 14~\mathrm{s}$ and $\delta t_{\phi=1/2} = z_{\mathrm{w},\phi=1/2}/c \approx 81~\mathrm{days}$ in the observer frame.
The steeper the density profile, the shorter the delay time becomes.
If gamma-rays are emitted from accelerated high energy electrons and positrons in the jets, the delay time of those gamma-ray emissions can discriminate the density stratification along the jet.
For examples, \textit{Fermi} Gamma-ray Burst Monitor (GBM) detected a gamma-ray-burst (GRB 170817A) with a time delay of $\sim 1.7~\mathrm{s}$ with respect to the merger time of the gravitational wave event (GW170817) \citep{Abbott:2017it}.
If this event is driven by a jet from a NDAF disk, such a short time interval is consistent with our models because the time interval between the gamma-ray emission and the gravitational wave emission is expected to be in the range between $\delta t_{\mathrm{obs},\phi=1} = \delta t_{\phi=1}\,(1 - \sqrt{\Gamma^2-1} \cos{\theta} / \Gamma) \approx 7.9\times 10^{-2}~\mathrm{s}$ and $\delta t_{\mathrm{obs},\phi=1/2} =  \delta t_{\phi=1/2}\,(1 - \sqrt{\Gamma^2-1} \cos{\theta} / \Gamma) \approx 11~\mathrm{hours}$ if the jet were directed at angle $\theta = 6~\mathrm{degrees}$ with respect to the line of sight to the observer located far away from the source.
Moreover, observations with the High Energy Stereoscopic System (H.E.S.S.) detected high-energy X-ray and Gamma-ray emissions $\sim 10~\mathrm{hours}$ after a prompt emission of GRB 180720B triggered by the \textit{Fermi} GBM \citep{Abdalla:2019dn}.
Therefore, the density gradient of those events is consistent with the range of the jet collimation profile we used here.
However, the opening angle of $45$ degrees for $\phi=1$ in the jets from NDAFs seems to be far from reality according to the many other observations of GRBs \citep{2018A&A...609A.112G}, in which a typical opening angle is assumed to be $5$ degrees.
Therefore, we may take into account hybrid collimation profiles, such as a two-component jet model \citep[the so-called spine-sheath jet model:][]{2003ApJ...594L..23V,2005NCimC..28..439P}.
First and foremost, we need to incorporate a model based on numerical experiments of the magnetically driven jets penetrating through the progenitor ambient medium and propagating at the further distance for more detailed investigations.
%

\subsection{Maximum energy of accelerated protons}
In this study we have estimated the maximum energy of accelerated protons by integrating the work done by wakefield force generated by a strong EM wave pulse over the maximum acceleration distance at the condition of $\omega_\mathrm{p}(z_\mathrm{w})/\omega_\mathrm{0} \lsim 1$.
This condition is based on the idealized assumption for the wakefield accelerations, such as a wakefield generated behind a short intense EM pulse.
In astrophysical wakefield acceleration, the EM wave pulse could be a broader and more complicated pulse structure.
If such a composite large amplitude Alfv\'enic wave pulse (which have turned into an EM wave pulse) occurs, it is expected to have incessant repetition of acceleration and out-of-acceleration within the excited wakefield.
For this reason, the spectrum for number of charged particles accelerated at the energy $\varepsilon$ has a power-law in $\varepsilon^{-p}$ with an index $p=2$ \citep{1991AIPC..230...27M}. 
In the wakefield acceleration in the jet, the magnetic field act as a guide for propagating EM wave pulses which excite the wake in plasma behind the pulses.
According to the dispersion relation of waves in plasma with no-magnetic field for simplicity, $\omega^{2} = \omega^{2}_\mathrm{p} + k^{2} c^{2}$, the group velocity of EM wave pulses, $v_\mathrm{g} = d\omega/dk = c^{2} / (\omega/k)$, is equal to the phase velocity of the wake, $v_\mathrm{p,w} = v_\mathrm{g} = c\sqrt{1-(\omega_\mathrm{p}/\omega)^{2}}$, thus the Lorentz factor of the wake becomes $\gamma_\mathrm{w} = 1/\sqrt{1- (v_\mathrm{p,w}/c)^{2}} = \omega / \omega_p$ where $\omega$ is the frequency of an EM wave pulse.
Once the wake phase velocity approaches nearly the speed of light beyond $z > z_\mathrm{w}$ (namely $\gamma_\mathrm{w} > 1$) along the jet, the electrons cannot compensate the immense wakes produced by EM wave pulses because the response of electrons is limited by the speed of light.
In other words, the EM wave pulses can "run away" from the instability due to the interaction between wave and plasma, and it can continue to "run though" plasma at the large distance, creating the large-scale coherent wakes in the jet.
Therefore the electric wakefields exert over the pump depletion length behind the EM wave pulses.
This is the reason why we consider that the wake has stability and rigidity under the condition of $\omega_\mathrm{p}/\omega <1$.
Unlike the Fermi acceleration, the wakefield acceleration do not require the multiple reflections by a magnetic mirror, which causes asynchronicity and the serious synchrotron radiation loss not only for electrons and positrons but also for protons with $\gsim 10^{20}~\mathrm{eV}$.
Since the wakefield force, not magnetic bending force, is responsible for accelerating charged particles, the wakefield acceleration is linear and synchronous with the propagating EM wave pulses and has more advantage in accelerating the super-EHECRs of $10^{22 - 23}~\mathrm{eV}$ in comparison with the Fermi acceleration, especially in the strongly magnetized jets from NDAFs.
Another point is that the accelerated high energy protons larger than the break energy, $\sim 10^{16}~\mathrm{eV}$, lose its energy by pion production through photo-meson interaction with energetic photons $\sim 1~\mathrm{MeV}$ in the jet \citep{1997PhRvL..78.2292W, 2001LNP...576.....L}.
Such energetic photons could be generated by the synchrotron radiation originated from accelerated electrons and positrons which are instantly generated by the wakefield acceleration in the jet less than picoseconds (the lowest range of color-scale in Figure~\ref{fig:max_energy}).
If the energy density of gamma-ray photons in the jet is larger than $10^{11}~\mathrm{erg/cm}^{3}$, the timescale of pion production is less than microseconds which is comparable to the acceleration time for energetic charged protons of $< 10^{20}~\mathrm{eV}$  (see Figure~\ref{fig:max_energy}).
This argument suggests that simultaneous production of high energy protons and reduction of its energy via pion production must be taken into account for evaluating the maximum energy of protons in more realistic condition.
Note that the high energy protons accelerated by the wakefield acceleration could be an alternative source of $\sim 10^{14}~\mathrm{eV}$ neutrinos as previously expected by \citet{1997PhRvL..78.2292W} in the context of a fireball model for GRBs \citep{1994AIPC..307..543P}.
Our model thus involves pinpointed spatial source of the emitter  as well as its temporal structure, from which we could diagnose the conditions of the emitter \citep{2019arXiv190805993C}.
This will be tested by estimating the flux of both gamma-ray emissions and high-energy neutrinos and comparing with the observations of IceCube \citep{2017ApJ...843..112A} in the context of our model in the near future.
%

\section{Conclusion}\label{sec:conclusion}
We have demonstrated the wakefield acceleration in the jets from NDAF disks as a model of gamma-ray bursts.
The wakefield acceleration postulates various observational signatures which could have been detected in the future.
The time-variability of $\lsim 100~\mathrm{MeV}$ neutrino emissions from the NDAF disks may discriminate the nature of an EM wave pulse which is responsible for driving the wakefield in the jets.
The tracing of gamma-ray emissions from high energy electrons and positions and subsequent burst of $\sim 10^{14}~\mathrm{eV}$ neutrinos may disclose the onset of wakefield acceleration in the jets.
The detection of the extremely high energy cosmic rays (EHECRs) of $10^{21 -22}~\mathrm{eV}$ and the super-EHCRs of $10^{22 -23}~\mathrm{eV}$ within several hours after both gamma-ray emissions and neutrino bursts could be a smoking gun for the astrophysical wakefield acceleration.
Because of those collective nature of high energy astrophysics and particles acceleration in ultra-relativistic regimes, the wakefield acceleration will be a key player for the multi-messenger astronomy.
%

\begin{acknowledgments}
The authors would like to thank Prof. R. Matsumoto, Prof. H. Sobel, Prof. S. Nagataki, and Prof. T. Totani for fruitful discussions, and Prof. K. Kohri for helpful comments.  We also thank the anonymous referee for constructive comments that help us to clarify our results and its limitation. This work is supported in part by the Norman Rostoker Fund.
\end{acknowledgments}

%







\bibliography{ms}{}
\bibliographystyle{aasjournal}



\end{document}